\documentclass[preprint,amsmath,showpacs]{revtex4}
\usepackage{graphicx}
\def\bee{\begin{eqnarray}}
\def\eee{\end{eqnarray}}

\begin{document}
\draft
\title{GeV to TeV astrophysical tau neutrinos}
\author{H. Athar$^{1,}$\footnote{E-mail: athar@phys.cts.nthu.edu.tw} and
C. S. Kim$^{1,2,}$\footnote{E-mail: cskim@yonsei.ac.kr}}
\address{$^{1}$  Physics Division, National Center for Theoretical Sciences,
Hsinchu 300, Taiwan\\
$^{2}$ Department of Physics, Yonsei University, Seoul 120-749, Korea}
\date{\today}
\begin{abstract}
Neutrinos with energy greater than GeV are copiously produced
in the $p(A,p)$ interactions occurring in several astrophysical
 sites such as (i) the earth atmosphere,
(ii) our galactic plane as well as in (iii) the galaxy clusters.
 A comparison of the tau and mu neutrino
flux in the presence of neutrino oscillations from
these three representative astrophysical sites is presented. It is pointed out
that the non-atmospheric tau neutrino flux starts dominating over the
 downward going atmospheric tau neutrino flux for neutrino energy $E$ as low as
  $\sim $ 10 GeV. This
 energy value is much lower than the energy value, $E \geq 5\times 10^{4}$ GeV, estimated
 for the dominance of the non-atmospheric mu neutrino flux,
 in the presence of neutrino oscillations. Future prospects for
possible observations of non-atmospheric tau neutrino flux are briefly mentioned.

\end{abstract}
\maketitle
\section{Introduction}
A present day main motivation for the extra-terrestrial neutrino astronomy
is to obtain the first evidence of tau neutrinos from the cosmos around us
 above the relatively
well known atmospheric neutrino background \cite{Athar:2002rr}. The
tau neutrinos are essentially an unavoidable consequence of almost maximal neutrino flavor
mixing between $\nu_{\mu}$ and $\nu_{\tau}$ as suggested by the
 atmospheric mu neutrino data analysis of the
 high statistics Super-Kamiokande detector
(SKK) \cite{Fukuda:2000np}.

A recent SKK analysis of the $L/E$ distribution of the
atmospheric mu neutrino data imply the following range of
neutrino mixing parameters \cite{Ashie:2004mr}
\bee
 1.9\times 10^{-3}\, \, \, {\rm eV}^{2}< \, \Delta m^{2} <3.0 \times 10^{-3}\, \, \,
{\rm eV}^{2}, \, \, \,
 \sin^{2}2\theta >0.9.
\label{range}
\eee
This is a  $90\% \, {\rm C.L.}$ range
with the best fit values approximately given by
$\Delta m^{2}=2.4\times 10^{-3}\, \, \, {\rm eV}^{2}$ and  $\sin^{2}2\theta =1$
 respectively.
 This range of neutrino mixing parameters results in purely
two flavor oscillation explanation of
the $L/E$ distribution  of the atmospheric
mu neutrino flux disappearance.
 The tau neutrinos as a result of these
 $\nu_{\mu}\to \nu_{\tau}$ oscillations are so far identified on
 statistical basis (rather than on event by event basis) \cite{icrc2003}.
  The total number of
 observed atmospheric non-tau neutrinos,
 on the other hand, are by now greater than $10^{4}$ from various
 detectors ranging in energy between $\sim 10^{-1}$ GeV to $\sim 10^{3}$ GeV \cite{Kajita:2000mr}.
 Given the recent detector developments, it is of some interest to estimate the tau neutrino flux from the
earth atmosphere as well as from the nearby astrophysical sites
 in order  to provide a
more complete basis for the hypothesis of $\nu_{\mu}\to \nu_{\tau}$
 oscillations.

    The expectations for high energy astrophysical mu neutrino
 flux with $E \geq 10^{3}$ GeV without neutrino
oscillation effects are summarized in  \cite{Gaisser:1997aw}.
 A purpose of this paper is to point out through
 examples that
the GeV to TeV (1 TeV = $10^{3}$ GeV) tau neutrino astronomy can be
quite {\tt different} as compared to the
below TeV mu neutrino astronomy which is
essentially dominated by the study of atmospheric
neutrinos only. The recently discovered neutrino
flavor oscillations does not bring any significant changes
as far as the below TeV extra-terrestrial mu neutrino search/astronomy is concerned.
 Considering a different neutrino flavor in the above energy range
may  change the
situation somewhat by opening a {\tt new} window to study cosmos.
In this context, we shall distinguish between the possibilities offered by the
different neutrino flavor astronomy
through some examples.
 For a summary of above TeV astrophysical tau neutrino fluxes,
 see \cite{Athar:2003mt}. In this
 letter, we shall focus on the tau neutrinos and estimate
the total tau neutrino flux from the
three different astrophysical sites in order to illustrate that
the GeV to TeV tau neutrino astronomy may be quite different
from the GeV to TeV mu  neutrino astronomy.

The neutrino oscillation probability in the two neutrino flavor
approximation is  \cite{Athar:2002uj}
\bee
 P(\nu_{\mu}\to \nu_{\tau})=\sin^{2}2\theta \sin^{2}
 \left(1.27 \frac{\Delta m^{2}({\rm eV^{2}})L({\rm km})}{E({\rm GeV})}\right).
 \label{osc-prob}
\eee
Here $L$, is the neutrino flight length. In the earth atmosphere, it
 can be estimated using
\bee
 L=\sqrt{(h^{2}+2R_{\oplus}h)+(R_{\oplus}\cos \xi)^{2}}-R_{\oplus} \cos \xi.
 \label{len}
\eee
The $L$ is essentially the distance between the detector and the height
 at which the atmospheric mu neutrinos are produced.
 The $R_{\oplus}\simeq 6.4 \cdot 10^{3}$ km is the earth radius, and  $h=15$ km
 is the mean altitude at which the atmospheric mu neutrinos are produced.
 In general, $h$ is not only a function of the zenith angle $\xi$, the neutrino flavor
 but also the neutrino energy \cite{Gaisser:1997eu}.

This letter is organized as follows. In section 2, the mu and tau neutrino flux
originating from the earth atmosphere,  the
galactic plane as well as from the galaxy clusters is briefly discussed.
In section 3, the neutrino oscillation effects
are studied for these. In section 4, the limited future prospects
for possible observations of non-atmospheric tau neutrinos are mentioned,
whereas in section 5, conclusions are presented.
\section{Some Examples of GeV to TeV astrophysical tau neutrinos}

    In order to estimate the total tau neutrino flux from a
given astrophysical site which is composed of intrinsic and oscillated ones,
 we need to first estimate the intrinsic mu and tau neutrino
flux from that site. In this section, we briefly describe the
general procedure used to estimate these.

    Before doing that, let us remark here that the incident cosmic ray
energy that we consider here ranges typically between
$10 < E_{p}/{\rm GeV} < 10^{4}$. The corresponding
c.m energy range in $pp$ collisions is $1 < \sqrt{s_{pp}}/{\rm GeV} < 10^{2}$ as
$s_{pp} \sim 2m_{p}E_{p}$. The data from various collider experiments indicates
that the tau neutrino parent hadron ($D^{\pm}_{S}$) production cross section here
is suppressed relative to mu
neutrino parent hadron  ($\pi^{\pm}$) production
cross section in $pp$ collisions in the above $\sqrt{s_{pp}}$
  range \cite{Hagiwara:fs}.
 Therefore, any sizeable tau neutrino
flux from the following sites should be verifying the neutrino
flavor mixing.
\subsection{Atmospheric}

Briefly, the incoming cosmic rays interact with the air nuclei $A$, in the
 earth atmosphere and give rise to mu neutrino flux.
For  $1 \leq E/{\rm GeV} \leq 10^{3}$, the $\pi^{\pm}$, $K$ production and
 their direct and indirect decays
 are the main sources of mu neutrinos, both being in the region of conventional
  mu neutrino production \cite{1962}. The absolute normalization
of the conventional atmospheric neutrino flux is presently known to be no better
than (20$-$25)\% \cite{Battistoni:1999at}.

For present estimates, the mu neutrino flux is taken from
 Ref.~\cite{Honda:1995hz}. These are
 neutrino flux calculations in one dimension without geomagnetic field
 effects. The up down
mu neutrino flux is taken to be the same, as the present discussion is
independent of any specific detector. At higher energy, the prompt mu
neutrino production from $D$'s dominates over the conventional one \cite{Volkova:gh}.

The atmospheric tau neutrino
flux arises mainly from $D^{\pm}_{S}$  and is
calculated in Ref.~\cite{Pasquali:1998xf,Athar:2001jw}.
 The Quark Gluon String Model (QGSM) is used in Ref.~\cite{Athar:2001jw} to
 model the $pA$ interactions.  The low energy
atmospheric tau neutrino flux is essentially isotropic \cite{Pasquali:1998xf}.
 For $E\leq  10^{3}$ GeV, the atmospheric tau neutrino flux is obtained by
 following the procedure given in Ref.~\cite{Pasquali:1998xf,Athar:2001jw}
 and re-scaling w.r.t new cosmic ray flux
spectrum, taking  it to be dominantly
the protons \cite{Gaisser:2002jj,ICRC2001,Honda:2004yz}.
\subsection{Galactic Plane}
The galactic plane mu neutrino flux
 is calculated in Ref.~\cite{Stecker:1978ah,Ingelman:1996md}, whereas
the galactic plane tau neutrino flux is calculated in Ref.~\cite{Athar:2001jw}.
 These calculations consider $pp$ interactions inside the galaxy with
 target proton number density $\sim $ 1/cm$^{3}$ along the galactic plane,
 under the assumption that the cosmic ray flux spectrum  in the Galaxy is constant
 at its locally observed  value.
 The current Energetic Gamma Ray Experiment Telescope (EGRET)
observations of the diffuse gamma-ray emission from the
galactic plane seems to imply a slightly less ($\sim 0.62$) target
proton density along the galactic plane \cite{Hunger:1997we}. We have checked that taking into account
this in our estimates makes only a minor difference.

Following Ref.~\cite{Athar:2001jw}, the galactic plane mu and tau
 neutrino flux for $E\leq  10^{3}$
 GeV is obtained by re-scaling w.r.t new cosmic ray flux
spectrum. The tau neutrino production is rather suppressed in the
galactic plane relative to mu neutrino production.
 The  mu neutrino flux is larger than the tau
neutrino flux for $E\leq  10^{3}$ GeV from the above two sites.

\begin{figure}
\includegraphics[width=7.in]{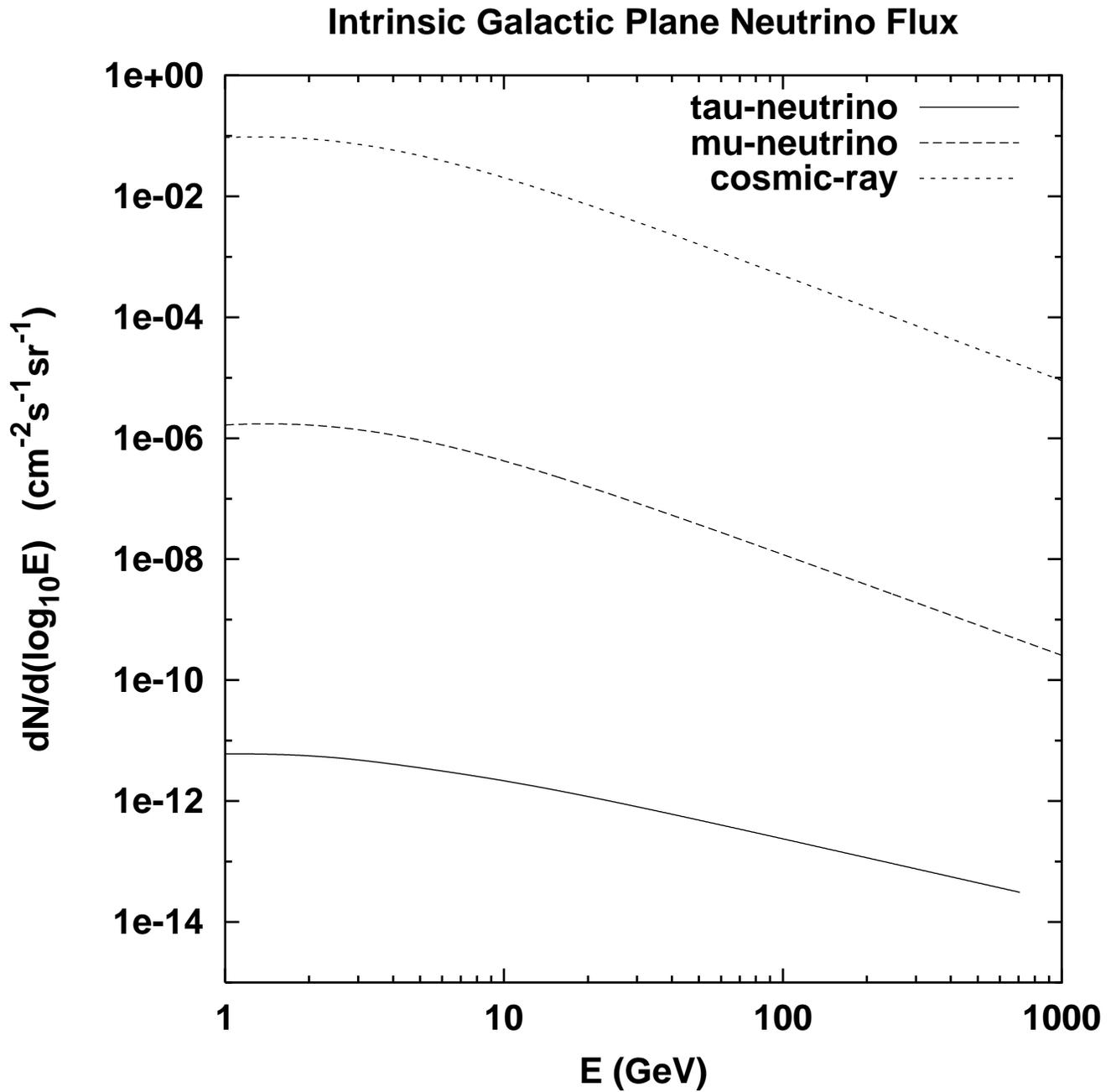}
\caption{\label{fig1}The intrinsic galactic plane
         mu and tau neutrino fluxes. The cosmic-ray
         flux spectrum used in the estimates is also shown.}
\end{figure}

Fig.\ref{fig1} gives the intrinsic mu and tau
neutrino flux,
 $F_{\nu}(E)\equiv {\rm d}N^{0}_{\nu}/{\rm d(log_{10}}E)$
 in units of cm$^{-2}$s$^{-1}$sr$^{-1}$, estimated using the above description. The
figure also shows the cosmic-ray proton flux spectrum
we have used.
\subsection{Clusters of galaxies}
    Clusters of galaxies are presently considered to be
an interesting laboratory to investigate various stages of
structure formation via the study of diffuse gamma-rays
(and cosmic-rays) from
 these \cite{Dar:1995tf,Berezinsky:1996wx}.

    The mu and tau neutrino fluxes are produced here
 in $pp$ interactions (of cosmic-rays with the intra cluster gas),
 under the main assumption that a large fraction of the
cosmologically produced baryons is inside these galaxy
 clusters. The intrinsic tau neutrino
flux estimate here is rather similar to the galactic plane
situation.

The maximum mu neutrino flux from clusters of
galaxies is estimated by correlating it with the corresponding
gamma-ray flux in $pp$ collisions.
 This estimate is under the assumption that diffuse gamma-ray flux
observed in the energy range 10 $ \leq E/{\rm GeV} \leq 50$
by EGRET is dominantly from clusters of galaxies. We shall use this upper limit
 as an another example of an astrophysical site that
may produce tau neutrino flux in the GeV to TeV energy
range.

The detailed model dependent calculations indicate that
the actual  contribution of the galaxy clusters
towards the recently observed extra-galactic diffuse gamma-ray
flux is approximately two orders of magnitude small \cite{Reimer:2003er}.

Before studying the effects of neutrino oscillations on these intrinsic neutrino fluxes,
 let us here emphasize that the non-atmospheric mu and tau neutrino fluxes can only be
considered as upper limits on these fluxes in light
of the existing inherent uncertainties in estimating these.
\section{Effects of neutrino oscillations}
We shall perform here the two neutrino flavor oscillation
analysis.   In the context of two neutrino
flavor oscillations, there are only two
neutrino mixing parameters. The mixing
angle $\theta $ and the $\Delta m^{2}(\equiv m^{2}_{2}-m^{2}_{1})$. There
are no matter effects for these two flavors.

In the two flavor approximation, the {\tt total} tau neutrino flux from
 an astrophysical site is given by
\bee
 F_{\nu_{\tau}}(E) =
 P_{\mu \tau}(E)\cdot F^{0}_{\nu_{\mu}}(E)  +
 P_{\mu \mu}(E)\cdot F^{0}_{\nu_{\tau}}(E) ,
\label{tot}
\eee
where $P_{\mu \tau}(E)\equiv P(\nu_{\mu}\to \nu_{\tau})$ is given by Eq. (\ref{osc-prob})
 and $P(\nu_{\mu}\to \nu_{\mu})=1-P(\nu_{\mu}\to \nu_{\tau})$.  The
 $F^{0}_{\nu}(E)$ is the
 intrinsic neutrino flux  in units of cm$^{-2}$s$^{-1}$sr$^{-1}$
 and is taken according to discussion in section 2.

\begin{figure}
\includegraphics[width=7.in]{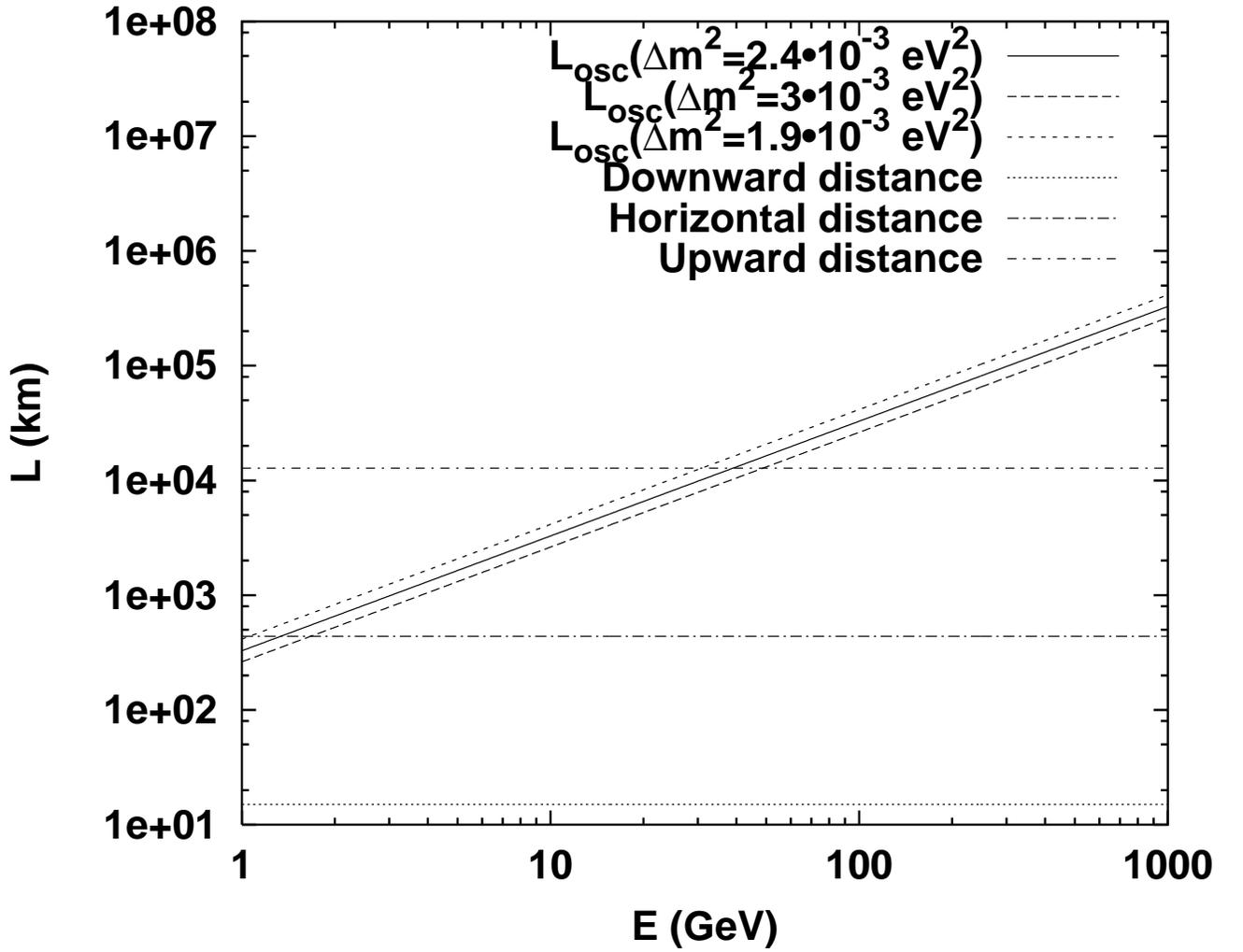}
\caption{\label{fig2}The $\nu_{\mu}\to \nu_{\tau}$
 oscillation length $L_{\rm osc}\, \, [\equiv 0.787 \, \, {\rm km} \, \, E({\rm GeV})/\Delta m^{2}({\rm eV}^{2})$]
 in km as a function of neutrino
 energy in GeV. The three general distances
 traversed by mu neutrinos in the earth atmosphere are also shown
 (the horizontal lines). More details are given in the text.}
\end{figure}

Three general directions in the earth atmosphere
are considered as representative examples to compare the atmospheric tau neutrino
flux with the non-atmospheric one in the {\tt presence} of neutrino oscillations.
 These are the downward, the horizontal
and the upward directions. Fig. \ref{fig2} depicts the $L_{\rm osc}$ given by
Eq. (\ref{osc-prob}) for the range of $\Delta m^{2}$ given by Eq. (\ref{range}) with maximal
 mixing.
 The three distances are taken from Eq. (\ref{len}), with, for instance,  the downward
distance is obtained by setting $\xi =0$. The horizontal distance is
obtained by setting $\xi = \pi/2$.

\begin{figure}
\includegraphics[width=7.in]{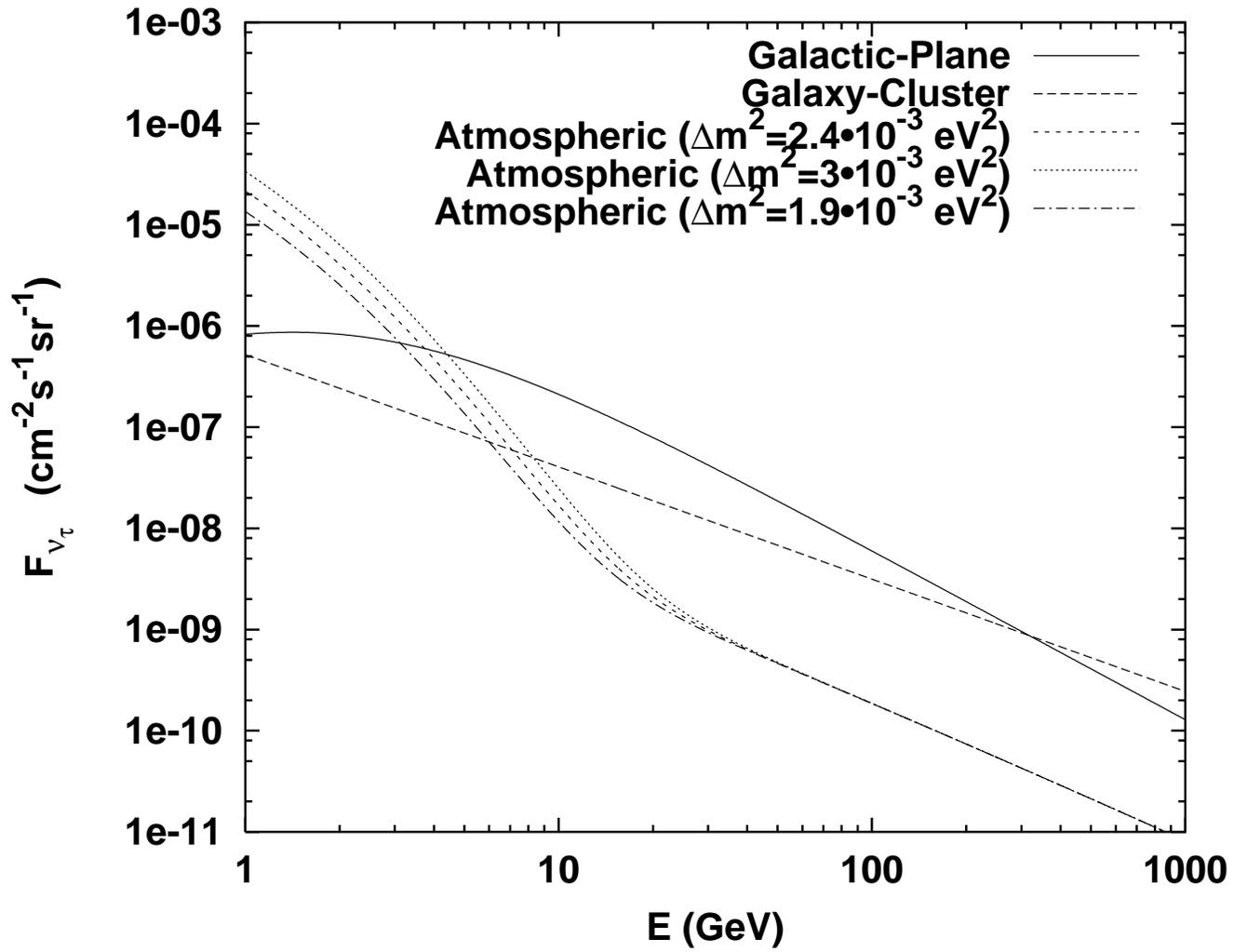}
\caption{An illustrative comparison of the galactic plane, the maximum galaxy clusters
  and the downward going atmospheric tau
 neutrino flux in the presence of neutrino oscillations as a function of neutrino
 energy.}
\label{fig3}
\end{figure}

Using Eq. (\ref{tot}), the total {\tt downward} going atmospheric
tau neutrino flux is estimated. It is then compared with the total galactic plane and
 total upper limit galaxy clusters
 tau neutrino flux in Fig. \ref{fig3} for the whole
range of $\Delta m^{2}$ with maximal mixing. The distance $L$ for galactic plane neutrinos
is taken as $\sim $ 5 kpc, where 1 pc $\sim  3\times 10^{13}$ km.
 For galaxy clusters, its representative value  is taken as 1 Mpc. Since
$L_{\rm osc} \ll L$, the galactic plane and galaxy cluster mu neutrinos oscillate
before reaching the earth. Also, note that this flux is averaged out
 for the whole range of $\Delta m^{2}$ in the entire considered energy range.
 The effect of different
$\Delta m^{2}$ values diminishes for $E \geq 50$ GeV for total
 atmospheric tau neutrino flux. From the figure, it can be seen
that the galactic plane/non-atmospheric tau neutrino flux starts {\tt dominating} over
the downward going atmospheric tau neutrino flux even for
$E$ as low as 10 GeV in the presence of neutrino oscillations.
 This is a very specific feature of {\tt tau neutrinos},
 and is absent for mu neutrinos. This specific
behavior has to do with the {\tt neutrino oscillations}. The
galactic plane tau neutrino flux for $1\leq E/{\rm GeV} \leq 10^{3}$ in the
presence of neutrino oscillations can be parameterized as
\bee
 F_{\nu_{\tau}}(E) =
 1.31\cdot 10^{-5} \cdot E^{1.07}\left[ E+2.15\exp({-0.21\sqrt{E}}) \right]^{-2.74},
\label{parameterize}
\eee
where $F_{\nu_{\tau}}$ is in units of
cm$^{-2}$s$^{-1}$sr$^{-1}$ and on r.h.s. $E$ is in units of GeV.

\begin{figure}
\includegraphics[width=7.in]{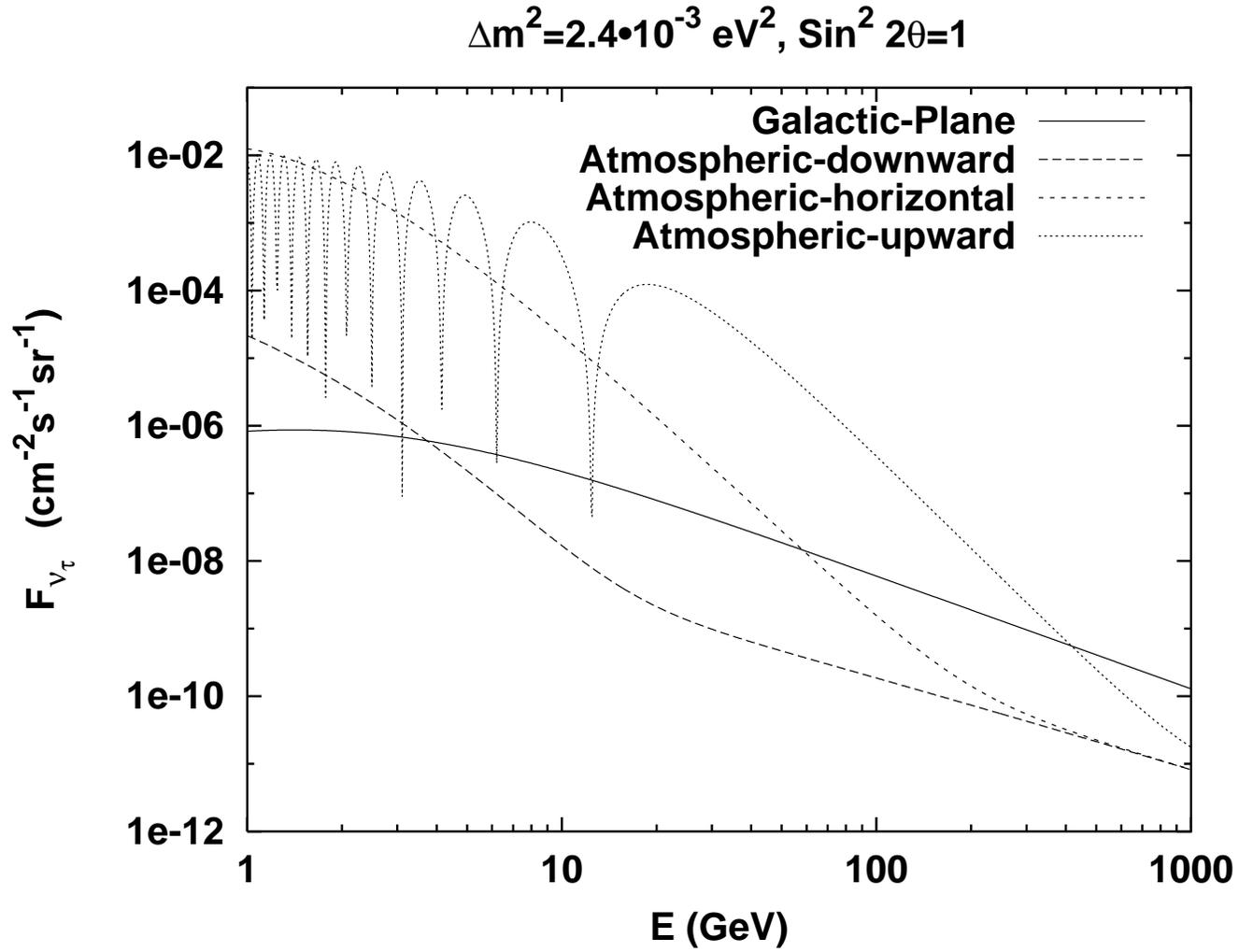}
\caption{The galactic plane and the atmospheric tau
 neutrino flux in the presence of neutrino oscillations for
 the three general directions in the earth atmosphere as a function of neutrino energy.
 The neutrino mixing parameter values used here are the approximate best fit values,
 $i.e.$, $\Delta m^{2}=2.4\times 10^{-3}$ eV$^{2}$ and $\sin^{2}2\theta =1$.}
\label{fig4}
\end{figure}

In Fig. \ref{fig4}, the galactic plane tau neutrino
flux is compared with the atmospheric tau neutrino flux,
 using Eq. (\ref{tot}) for the three general
directions for the atmospheric tau neutrino flux reaching the detector.
 Here the best fit values of the neutrino
mixing parameters are used.  The oscillatory nature of the upward
going tau neutrino flux can be seen from Eq. (\ref{osc-prob}).
The cross over for the galactic tau neutrinos relative to the
 horizontal atmospheric tau neutrinos
occurs at $\sim $ 50 GeV, whereas the same occurs for the upward
direction at $\sim $ 400 GeV. The total atmospheric tau neutrino flux is maximum
in the upward direction. It is
{\tt minimum} in downward direction, relative to the galactic plane
tau neutrino flux in the presence of neutrino oscillations,
 owing to the behavior of $L/L_{\rm osc}$ ratio as a function of neutrino energy.
 Fig. \ref{fig4} indicates that zenith angle
dependence of the total tau neutrino flux can at least in
principle help to distinguish between atmospheric and
 non-atmospheric tau neutrino flux.  The galactic tau
neutrino flux transverse to the galactic plane is
three to four orders of magnitude smaller than the
galactic plane one \cite{Athar:2001jw}.

\begin{figure}
\includegraphics[width=7.in]{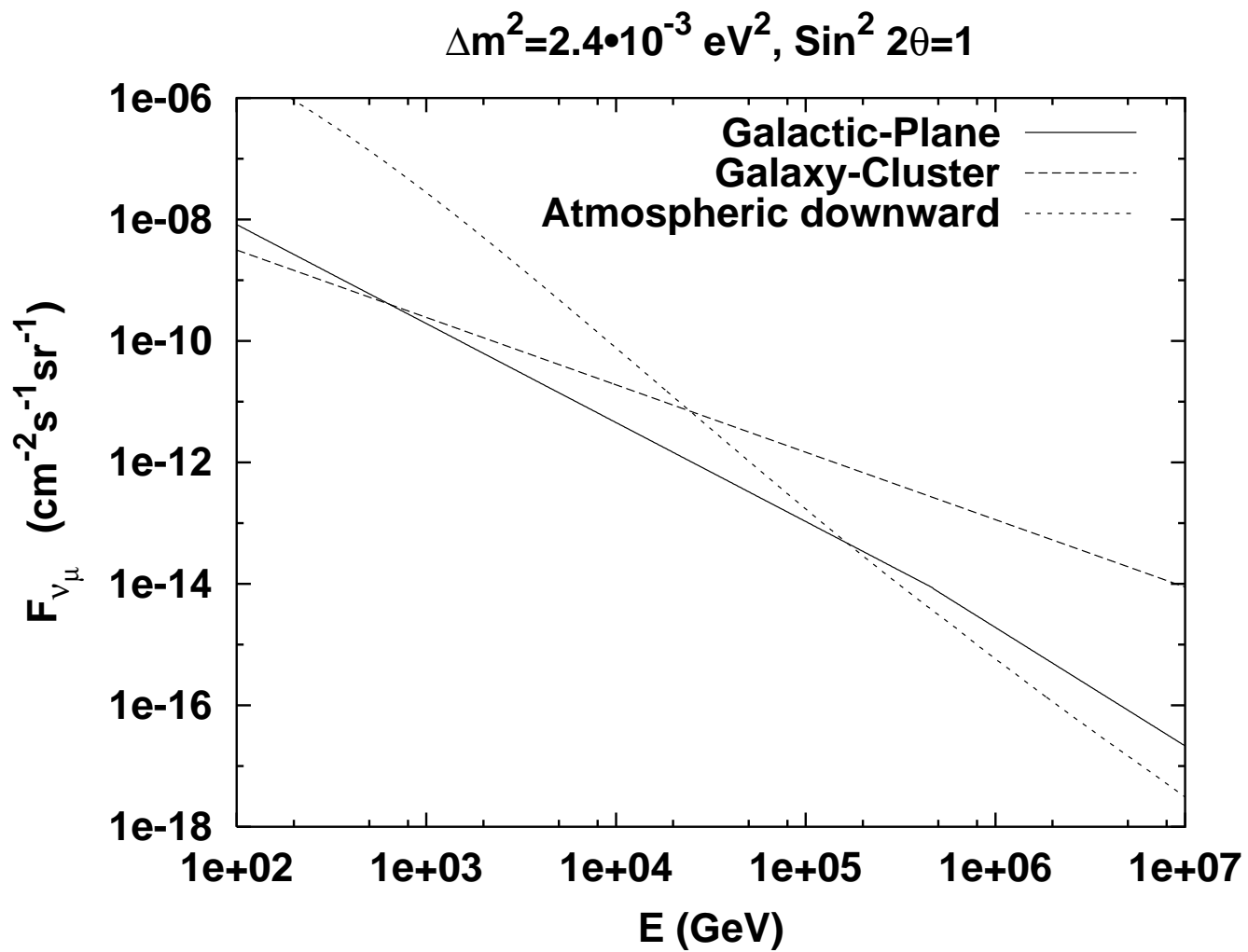}
\caption{Comparison of the downward going atmospheric mu neutrino
 flux, the galactic plane mu neutrino flux and the maximum galaxy
 cluster mu neutrino flux in the
 presence of neutrino oscillations. The non-atmospheric
 mu neutrino flux starts dominating over the atmospheric one
 only for $E \geq 5 \times 10^{4}$ GeV.}
\label{fig5}
\end{figure}

Fig. \ref{fig5} gives a comparison of the downward going atmospheric and
the galactic plane mu neutrino flux in the presence of neutrino
 oscillations. For this comparison, mu neutrino
 flux is taken from Ref.~\cite{Gondolo:1995fq} without re-scaling
 for $E \leq 10^{3}$ GeV. This
mu neutrino flux includes contribution from the $D$'s
 for $E \geq 6.3\times 10^{5}$ GeV.
The total mu neutrino flux is
 estimated according to Eq. (\ref{tot}) with appropriate
 modifications for the best fit values of the two neutrino mixing parameters.
 In contrast to the
possibility of seeing the galactic plane with multi GeV
tau neutrinos, note here that with mu neutrinos, it
can occur only for $E \geq 10^{5}$ GeV.

A relevant remark is that for the best fit values of the neutrino mixing parameters,
the $P(\nu_{\mu}\to \nu_{\tau})$ is relatively large along the
horizontal and upward directions in the earth atmosphere
 [see Eq. (\ref{osc-prob}) and Fig. \ref{fig1}] for $1 \leq E/{\rm GeV} \leq 10$.
 So essentially the atmospheric mu neutrino flux in the absence of
neutrino oscillations alone determine the total atmospheric
tau neutrino flux in comparison with the total galactic
 tau neutrino flux.

\section{Prospects for possible future observations}
In this section, we shall estimate the galactic plane tau
neutrino induced shower event rate for a  megaton class
of detectors to indicate the limited prospects offered
by GeV to TeV tau neutrino astronomy to search for
extra atmospheric astrophysical neutrino sources in this energy range.

    Let us add a remark here that, at present, the  dedicated large high energy neutrino detectors
 such as the Antarctic Muon And Neutrino detector Array
(AMANDA) are also sensitive to all neutrino flavors, however,  with energy
 $ \simeq 50 $ TeV \cite{Ackermann:2004zw}.

For $10 \leq E/{\rm GeV} \leq 10^{3}$, a signature for the tau neutrinos is to
 measure the energy spectrum of the tau lepton induced electromagnetic and hadronic
 showers
produced in tau neutrino nucleon interactions
occurring inside a densely instrumented Cherenkov radiation detector \cite{Stanev:1999ki}.
Though, it is a challenging task to distinguish between tau and non-tau
neutrinos for the present generation of detectors in the above
 energy range \cite{Hall:1998ey}, however certain shower signatures remain
distinctive  for tau neutrinos \cite{Stanev:1999ki}.

The galactic tau neutrino induced shower production  rate can be
 approximately estimated  by convolving the
total galactic tau neutrino flux in the presence of neutrino oscillations, given by
 Eq. (\ref{parameterize}) with the $\sigma^{\rm CC}_{\nu_{\tau}N}(E)$. The
$\sigma^{\rm CC}_{\nu_{\tau}N}(E)$ for $10 \leq E/{\rm GeV} \leq 10^{3}$ is parameterized
 as
\bee
 \sigma^{\rm CC}_{\nu_{\tau}N}(E)=-4.43+0.52\cdot E+3.58\cdot 10^{-4}\cdot E^{2}.
\eee
The $\sigma^{\rm CC}_{\nu_{\tau}N}$ is in units of $10^{-38}$ cm$^{2}$ and $E$
is in units of GeV. The CTEQ6 parton distribution functions \cite{Kretzer:2003it}
 were used to estimate the cross section.

For recent detailed evaluations of
 $\sigma^{\rm CC}_{\nu_{\tau}N}$, see Ref.~\cite{Paschos:2001np}.
 The possible tau lepton polarization effects \cite{Hagiwara:2003di}
 are not taken into account in the
 event rate estimates presented here.

  Table I
gives the galactic tau neutrino induced shower event rate for a one Mega ton detector,
 in units of $({\rm Mt}\cdot {\rm yr})^{-1}$
 in $2\pi $ steradians of upper hemisphere
 in eight logarithmically equally spaced energy bins.  The table indicates
 that the per year event rate is about 1.5.
 With a 3 to 5 year data collection time for a one Mega ton
detector, the galactic tau neutrino induced shower event rate can thus be in the range of
 $\sim \, {\cal O} (10)$ for $E \geq 10$ GeV.
 This detector faces only the downward going atmospheric tau neutrino
 flux as background to the dominant galactic plane tau neutrino
 flux in the presence of neutrino oscillations.

\section{Conclusions}

1. The effects of neutrino oscillations on
low energy (1 GeV $\leq E \leq 10^{3}$ GeV) tau neutrino flux produced in the earth atmosphere,
in our galactic plane and in galaxy clusters  are presented in two neutrino flavor approximation.

2. The galactic plane should be observable
with tau neutrinos with energy $\geq 10$ GeV, depending on the orientation
of the concerned detector w.r.t. galactic center/plane at the time of observation.
The observation of galactic plane with multi GeV {\tt tau neutrinos} is in sharp
 contrast to the case
of mu neutrinos with which the galactic plane is observable only with energy
$\geq  10^{5}$ GeV for the same orientation of the detector.

3. Transverse to the galactic plane, the maximum galaxy clusters
tau neutrino flux is also above the atmospheric one, providing
another example of a new window to study cosmos in the above energy
range.

4. This observation may also have some relevance
for the long baseline experiments searching for the tau neutrinos in
$\nu_{\mu}\to \nu_{\tau}$ oscillations \cite{lbl}.
\section*{Acknowledgements}
The work of H.A. is supported by  Physics Division of NCTS.
The work of C.S.K. was supported in part by CHEP-SRC
Program, in part by Grant No. R02-2003-000-10050-0 from BRP of the
KOSEF. H.A. would like to express thanks to IPAP, Yonsei University, where
the part of the work has been completed.
\pagebreak
\begin{table}
\caption{Galactic plane tau neutrino induced shower event rate for a Megaton class
 of detectors in eight logarithmically equal energy bins. Details are given in the text.}
\vspace{0.25in}
\begin{tabular}{|c|c|c|c|}
\hline
\hline
\label{tableone}
 {\em Energy Bin} (GeV) & {\em Event Rate} $({\rm Mt}\cdot {\rm yr} \cdot 2\pi {\rm \, sr})^{-1}$
 & {\em Energy Bin} (GeV) & {\em Event Rate} $({\rm Mt}\cdot {\rm yr} \cdot 2\pi {\rm \, sr})^{-1}$\\
 \cline{1-4}
  10    $-$ 17.78     &  0.22  & 100 $-$ 177.83   & 0.16    \\
  17.78 $-$ 31.62  &  0.30     & 177.83$-$ 316.23 & 0.12    \\
  31.62 $-$ 56.23  &  0.26     & 316.23$-$ 562.34 & 0.10    \\
  56.23 $-$ 100    &  0.22     & 562.34$-$1000    & 0.07    \\
\hline
\hline
\end{tabular}
\end{table}
\pagebreak
\pagebreak
\pagebreak
\pagebreak
\pagebreak
\end{document}